\documentclass{article}

\usepackage[
    top=1in,
    bottom=1in,
    left=1.0in,
    right=1.0in
]{geometry}

\usepackage{graphicx}
\usepackage{amsmath}
\usepackage{xcolor}
\usepackage{booktabs}
\usepackage{siunitx}

\title{\bfseries Assessing Risks of Hydro-Generator Shaft Fatigue from Data Center Load Oscillations}

\author{
Kaustav Chatterjee\footnote{\text{Corresponding author: kaustav.chatterjee@pnnl.gov}}, Meghana Ramesh, Shuchismita Biswas, Brett A. Ross,\\
Antos C. Varghese, Sameer Nekkalapu, Slaven Kincic\\[0.5em]
\small{\text{Pacific Northwest National Laboratory\footnote{The Pacific
Northwest National Laboratory is operated for the U.S. Department of Energy by Battelle Memorial Institute under Contract DE-AC0576RL01830}}}\\
\small Richland, Washington 99354, United States of America
}

\date{}

\begin{document}

\maketitle

\renewcommand{\abstractname}{Summary}

\begin{abstract}
Large AI data center loads can introduce persistent sub-synchronous active-power oscillations that may impact nearby generators by exciting torsional modes and increasing shaft stress. This paper presents a model-based framework for evaluating hydro-generator shaft fatigue risk under such oscillatory load conditions. The methodology is demonstrated using an electromagnetic transient simulation model consisting of a hydroelectric generator represented by a two-mass turbine-generator model with parameters derived from real-world generation units, and a configurable AI data center load.

The risk assessment is conducted in two stages. First, a network transfer function quantifies how active-power oscillations propagate from the data center point of interconnection to the hydro-generator terminal. A plant transfer function then captures the relationship between generator-side oscillations and the resulting shaft torque pulsations. A frequency-scan approach is used to isolate the response at individual forcing frequencies and identify resonance regions where shaft torque amplification is greatest. Parametric studies demonstrate that shaft torque amplification risk is strongly influenced by shaft-system characteristics, particularly the generator-to-turbine inertia ratio and torsional damping. Lower inertia ratios shift the dominant torsional mode to lower frequencies and can increase torque amplification, suggesting that some Kaplan-type configurations may be more vulnerable than comparable Francis or Pelton units, depending on shaft stiffness and damping. Lower damping significantly amplifies the resonant response and consequently increases fatigue exposure.

In the second stage, a simplified fatigue assessment methodology based on S--N curve principles and the Goodman diagram is introduced to relate dynamic simulation results to mechanical integrity. The resulting Goodman safety factor provides a practical fatigue-risk metric that can help grid planners and plant owners understand the impact of persistent AI data center load oscillations on the mechanical service life of hydroelectric plants. This approach supports interconnection studies, informs load oscillation limit requirements, and guides plant-level monitoring and mitigation strategies.
\end{abstract}

\vspace{0.3em}

\begin{quotation}
\small

\noindent\rule{\linewidth}{0.5pt}

\vspace{0.4em}


\vspace{0.4em}

\noindent\rule{\linewidth}{0.5pt}

\end{quotation}

\vspace{0.1em}

\section{Introduction}

The United States (US) is experiencing a rapid increase in electricity demand, driven largely by hyper-scale data centers supporting artificial intelligence (AI) training and inference workloads \cite{ieee_spectrum_datacenter_boom_2026}. A recent report projects that data centers could account for 6.7–12\% of total US electricity demand by 2028 – a significant increase from approximately 4.4\% in 2023 \cite{shehabi2024datacenter}. This growth is intensifying resource adequacy and infrastructure expansion challenges. To accommodate load growth within existing infrastructure constraints and interconnection bottlenecks, two primary strategies are being pursued: co-locating new generation with data centers, or siting data centers near existing generation and high-capacity transmission infrastructure. The latter approach is particularly attractive because it leverages existing grid infrastructure to serve demand at relatively low cost. 

\begin{figure}[h]
    \centering
    \includegraphics[width=0.9\linewidth]{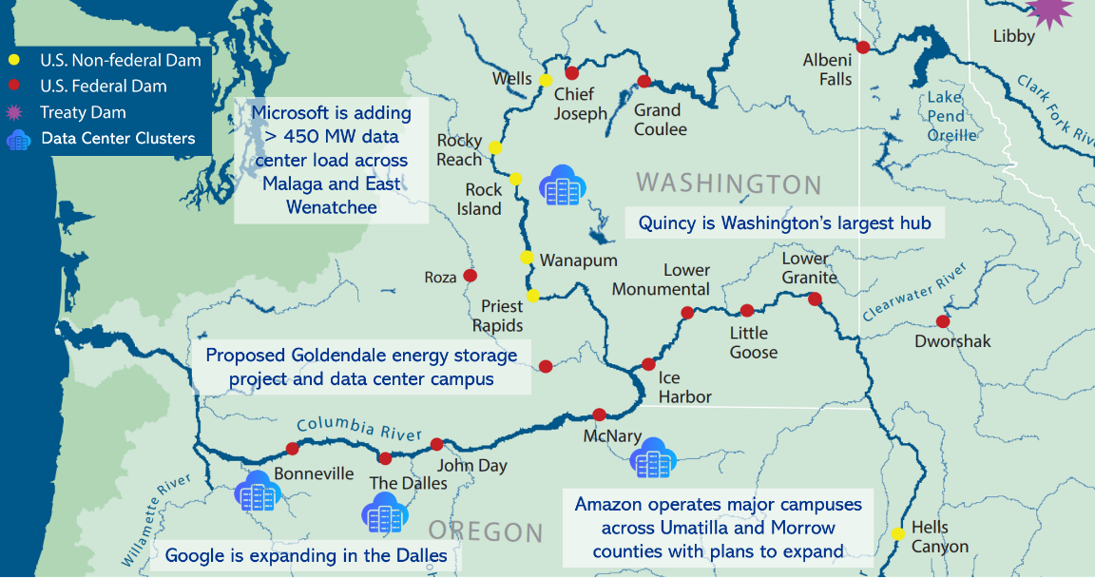}
    \caption{Data center growth along the Columbia River Basin in U.S. Pacific Northwest; data center locations are approximate. Basemap from  \cite{bpa_columbia_map}}
    \label{fig:columbia_map}
\end{figure}

Among existing generation resources, hydropower plants are emerging as strategically important assets supporting hyperscale facilities, due to their fast-ramping capability, frequency support services, and access to cooling water. In the Pacific Northwest, several data center clusters have already emerged near major hydropower plants along the Columbia River Basin \cite{katz2009techtitans} (Fig.~\ref{fig:columbia_map}), with similar trends emerging across the US \cite{webber2025hydrodatacenters}. Recent hydropower-data center agreements, including Google’s Hydro Framework Agreement with Brookfield Renewable for Susquehanna River facilities in Pennsylvania \cite{umble2025googlehydro}, further indicate that hydropower-backed data center siting may become a replicable development model. Globally, hydropower rich countries like Norway are also attracting major data center investments, with Microsoft and OpenAI citing hydropower availability as a key factor \cite{welsch2025norwayai}.

This surging demand warrants careful evaluation of the impact of data center operations on hydro assets. AI facilities exhibit tightly synchronized and rapidly varying load characteristics that can drive forced oscillations and excite sub-synchronous control and torsional interactions. This risk of oscillatory interactions had been identified as a high-priority concern by the NERC Large Loads Working Group (LLWG) \cite{nerc2025large_loads}. Recent studies have examined how torsional oscillations due to AI load fluctuations can cause shaft stress, fatigue accumulation, and loss of operating life in steam and gas-fired turbine generators \cite{zhang2026lel_sso, hossain2026ai_fatigue}. However, hydroelectric generation systems have received comparatively little attention despite their increasing proximity to hyperscale loads. This gap is important because hydro-generators have distinct governors, turbines, and shaft configurations that may respond differently to sustained oscillatory forcing than thermal units \cite{chatterjee_hydro_Access}. One reason hydroelectric units have received less attention in the context of torsional interactions is their lower susceptibility to classical sub-synchronous resonance (SSR) phenomena. Historical SSR events associated with series-compensated transmission systems and HVDC controls primarily involved steam turbine-generator units, where multiple turbine stages are connected through flexible shaft sections. These configurations can produce several torsional modes, some of which may overlap with electrical network resonance frequencies and lead to adverse electromechanical interactions \cite{epri2006torsional}. In contrast, conventional hydroelectric units generally have larger rotating masses, slower shaft dynamics, higher generator-to-turbine inertia ratios, and lower dominant torsional frequencies. These characteristics have traditionally provided stronger damping and reduced the likelihood of adverse torsional interactions under classical network-induced SSR conditions \cite{bladh2013torsional, kundur1994power, eilts1979shaft}.

However, unlike classical SSR, AI workloads lead to persistent forcing components across a broad sub-synchronous frequency range, increasing the likelihood of resonance with hydro-generator torsional modes. This concern is heightened by modern hydro-generator designs and refurbishment practices that may reduce generator inertia through lighter rotor structures or replacement of older rotor poles  \cite{bladh2013torsional, eilts1979shaft}. Lower generator inertia reduces generator-to-turbine inertia ratios, that can in turn decrease modal damping and increase susceptibility to torsional amplification under sustained oscillatory loading \cite{andersson1984inertia}. As hyperscale data center loads proliferate near major hydropower resources, it becomes increasingly important to evaluate whether AI-driven demand variability could excite hydro generator torsional modes and impact mechanical integrity and stability.

To this end, this paper develops a model-based framework for assessing hydro-generator shaft fatigue risk from AI-training-induced forced oscillations, combining network-level propagation analysis with plant-level torsional response and a simplified fatigue interpretation. The framework is demonstrated using electromagnetic transient (EMT) simulations on representative models of real-world hydro-generation units. Through the outlined methodology, the paper identifies hydro-generator configurations and operating conditions under which persistent AI-load oscillations may pose risks of increased equipment fatigue. Analysis shows that although hydro-generators have higher torsional damping than steam and natural gas units, shaft oscillation amplification can still be significant for designs with low generator-to-turbine inertia ratio and torsional frequency. Units with Kaplan turbines tend to have lower generator-to-turbine inertia ratios and hence may have higher susceptibility, although the shaft's torsional spring constant and damping also impact risk. Since viscous damping in hydraulic turbines is often difficult to estimate, risks of torsional amplification may be underestimated — warranting careful risk assessment across a range of expected values to account for model parameter uncertainty. 

\section{Problem Description}
Large AI training loads can produce tightly synchronized, rapidly varying demand patterns that act as persistent sources of forced oscillations in power systems. These aggregate demand profiles are multi-periodic and may contain oscillatory components across several frequency ranges \cite{choukse2025power}. Lower-frequency components in the 0.1–1 Hz range can propagate over wide areas of the transmission network and influence bulk-system dynamic behavior \cite{biswas2025large_load}. In contrast, higher-frequency components greater than 5 Hz tend to remain localized, can cause sub-synchronous control and torsional interactions, and are therefore of particular concern for generators near data centers \cite{choukse2025power}.

This work focuses on the impact of the high-frequency components of data center oscillations ($>$ 5 Hz) on neighboring hydroelectric generating units. At these frequencies, active power oscillations at the generator terminals can excite torsional responses within the turbine-generator shaft system. If the forcing frequencies are close to the natural torsional modes of the turbine-generator assembly, the resulting shaft torque oscillations may become amplified, leading to elevated shaft stress, fatigue accumulation, and reduction in equipment operating life under sustained oscillatory conditions. Hence, understanding the frequency-dependent torsional response of hydro-generator systems is important for assessing the mechanical and operational risks associated with nearby hyperscale sites. These insights can help grid operators understand if performance standards (e.g. oscillation amplitude limits, frequency ranges to be avoided, etc.) should be imposed on load facilities, and hydro-plant owners decide if plant maintenance schedules need to be reevaluated to account for data center proximity.

The risk assessment methodology presented in this paper is organized around four questions: (a) which sub-synchronous forcing frequencies are most likely to amplify hydro-generator shaft torque; (b) how electrical distance and network propagation affect the oscillation magnitude reaching the generator; (c) which hydro-generator parameters (e.g., generator-to-turbine inertia ratio, turbine configuration, shaft stiffness, and effective damping) most strongly influence torsional susceptibility; and (d) how shaft torque pulsation amplitudes can be interpreted in terms of fatigue risk.

The remainder of the paper illustrates the proposed workflow using EMT simulations on a test system. The test system comprises a hydroelectric generator connected to a data center load and the larger grid modeled and is described in Section 3. By varying the generator parameters, the behavior of three representative hydro plants is studied. Section 4 uses outputs from the simulated model to define a transfer-function-based risk framework that separates network propagation from plant-level torsional amplification and identifies the forcing frequencies that produce the largest shaft torque response in the studied units. In Section 5, the torque-amplification metric obtained is applied in sensitivity studies to evaluate how different hydro-generator parameters affect oscillation susceptibility. The dynamic simulation results are translated to a simplified shaft fatigue metric in Section 6. Section 7 outlines key takeaways from this work and outlines future research directions.

\section{EMT Model Development} 
\begin{figure}[h]
    \centering
    \includegraphics[width=0.85\linewidth]{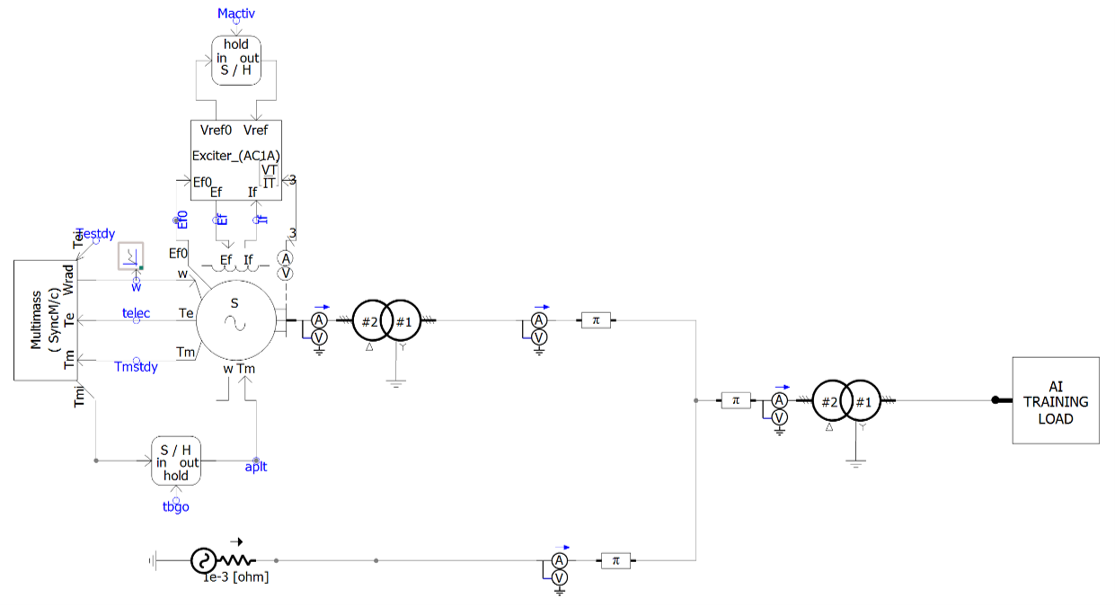}
    \caption{Screenshot from PSCAD showing the EMT model of the test system with hydro-generator and AI data center load.}
    \label{fig:emt_model}
\end{figure}

An EMT simulation framework was developed in PSCAD to evaluate the interaction between AI-load-induced forced oscillations and hydroelectric generation systems. The test system modeled consists of a hydroelectric generating unit connected through a transmission network to a configurable AI training load model capable of reproducing oscillatory demand behavior (see Fig.~\ref{fig:emt_model}). The PSCAD model is publicly available on GitHub \cite{pnnl_hydro_oscillations}. The individual components modeled are described next.

\textbf{\textit{Hydroelectric Generator Model}}: The generator is modeled with a salient-pole rotor configuration, typical of hydro units. The generator’s electrical parameters are obtained from \cite{bladh2013torsional}. A two-mass turbine-generator shaft model is used to capture the torsional dynamics between the turbine and generator masses. This representation enables the direct evaluation of torsional mode excitation and shaft torque oscillations due to simulated forced oscillations. By adjusting key parameters such as generator and turbine inertia constants (and therefore, generator-to-turbine inertia ratio), shaft stiffness, damping coefficients, turbine type, and generator ratings – the impact on different representative hydro units can be estimated. In this work – three different parameter sets have been utilized, as shown in Table ~\ref{tab:shaft_parameters}. Unit A describes a hydro generator from the Nordic grid modeled by tuning the IEEE benchmark model for sub-synchronous resonance \cite{bladh2013torsional}, while GC-7 and GC-19 are based on two units at the Grand Coulee power plant \cite{eilts1979shaft}. It should be noted that the GC-7 and GC-19 shaft parameters are drawn from a 1979 study \cite{eilts1979shaft} by the US Bureau of Reclamation and may not reflect the current as-built condition of these units. As discussed in Section 1, refurbishments such as rewinding, pole replacement, or use of lighter rotor structures can change the shaft assembly model parameters relative to their original design values. Where such modifications have occurred, the actual torsional frequency and amplification susceptibility of these units may differ from the values used here. The results in this paper should therefore be interpreted as illustrative of the proposed methodology rather than as a current condition assessment of these specific units.
While PSCAD supports detailed governor models, governor dynamics were not analyzed in this work because the frequency range of torsional modes is significantly higher than that of governor dynamics.

\begin{table}[h]
\centering
\caption{PSCAD multi-mass torsional shaft model parameters used for analysis.}
\label{tab:shaft_parameters}
\renewcommand{\arraystretch}{1.15}
\begin{tabular}{lccc}
\toprule
\textbf{Design Parameters} & \textbf{Unit A} & \textbf{GC-7} & \textbf{GC-19} \\
\midrule
Generator rating &
500 MVA & 60 MVA & 715 MVA \\

Rated speed &
150 rpm & 200 rpm & 72 rpm \\

Turbine inertia &
$1.68 \times 10^{6}$ kg m$^{2}$ &
$90.6 \times 10^{3}$ kg m$^{2}$ &
$8428 \times 10^{3}$ kg m$^{2}$ \\

Generator inertia &
$33.75 \times 10^{6}$ kg m$^{2}$ &
$0.653 \times 10^{6}$ kg m$^{2}$ &
$107 \times 10^{6}$ kg m$^{2}$ \\

Shaft spring constant &
$18.66 \times 10^{9}$ Nm/rad &
$0.208 \times 10^{9}$ Nm/rad &
$12.09 \times 10^{9}$ Nm/rad \\

Turbine damping &
$5.01 \times 10^{6}$ Nm s/rad &
$1.37 \times 10^{5}$ Nm s/rad &
$12.6 \times 10^{6}$ Nm s/rad \\

\midrule
\textbf{Derived Quantities} & \textbf{Unit A} & \textbf{GC-7} & \textbf{GC-19} \\
\midrule

Generator-to-turbine inertia ratio &
20.1 & 7.2 & 12.7 \\

Torsional frequency &
17 Hz & 8.14 Hz & 6.26 Hz \\

\bottomrule
\end{tabular}
\end{table}

\begin{figure}[h]
    \centering
    \includegraphics[width=0.95\linewidth]{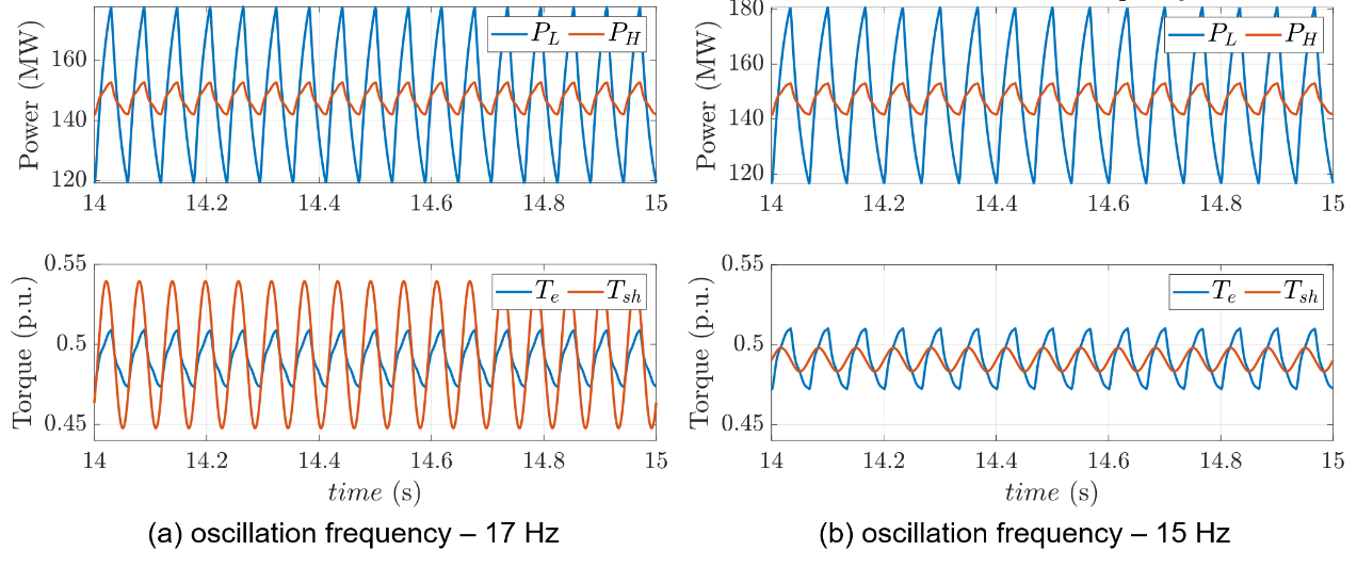}
    \caption{The impact of load oscillations on generator active power outputs, electrical torque, and shaft torque, simulated for Unit A. When the forcing frequency coincides with the generator torsional mode at 17 Hz, shaft torque oscillations are higher than electric torque. This amplification is not observed at 15 Hz.}
    \label{fig:torque_comp}
\end{figure}

\par\textbf{\textit{AI-training Data Center Load Model}}: The AI training load model used leverages the Data Center Model Library (DML) developed by the Pacific Northwest National Laboratory to represent the aggregated behavior of data centers in grid-level studies \cite{ross2026datacenter}. The model was configured to reproduce periodic active-power demand variations representative of synchronized GPU-based AI training workloads. It supports both replay of representative AI training load profiles and controlled injection of forced oscillations over selected frequency ranges. Key oscillation characteristics, including forcing frequency, amplitude, duty cycle, and synchronization pattern, can be adjusted to emulate different AI workload operating conditions. This configurable representation enables controlled evaluation of how load-induced oscillations propagate through the transmission network and affects nearby hydroelectric generating units. It allows the same hydro-generator model to be tested under prescribed sub-synchronous forcing conditions, making it possible to isolate frequency-dependent torsional response, resonance behavior, and shaft torque amplification. \\

\textbf{\textit{Transmission Network Model}}: A coupled $\pi$-section model of a short transmission line was adopted for interconnecting the data center load to the hydro-generator plant and the larger grid. Since the high-frequency oscillation phenomena investigated in this study are highly localized and the primary focus is the hydro-generator's torsional response, the bulk power grid is represented as an infinite bus. 

Figure~\ref{fig:torque_comp} shows the response of Unit A to simulated load oscillations at two forcing frequencies. The load active power at the POI is denoted by $P_L$, while the generator-side response is recorded as hydro-generator active power $P_H$, electromagnetic torque $T_e$, and shaft torque $T_{sh}$. The results show that shaft torque pulsations are strongly frequency dependent and become largest when the forcing frequency coincides with the unit’s torsional mode. The simulation outputs, particularly $T_{sh}$, $T_e$, and $P_H$, are used in the risk analysis framework described in the following section.

\section{Shaft Torque Amplification Risk Analysis}

\begin{figure}[h]
    \centering
    \includegraphics[width=\linewidth]{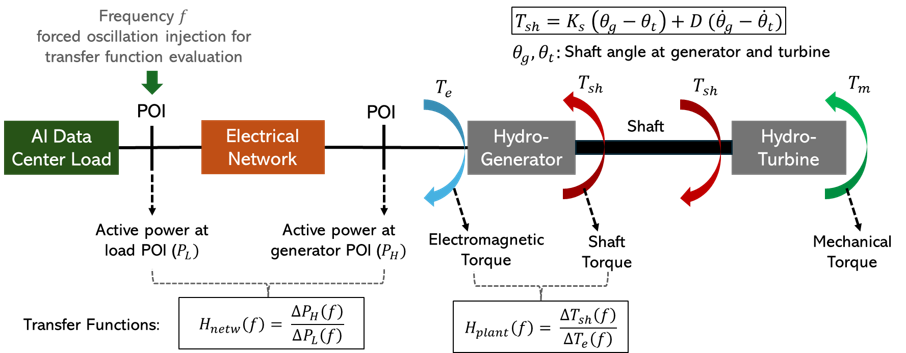}
    \caption{Analyzing how load-induced oscillations are translated to shaft torque pulsations.}
    \label{fig:shaft_analysis}
\end{figure}

The first step in analyzing the impact of forced oscillations on generator fatigue involves estimating how load-induced oscillations are reflected in shaft torque pulsations. This analysis proceeds in two parts, as shown in Fig.~\ref{fig:shaft_analysis}.

First, the transmission-network propagation of oscillations from the data center to the generating unit is evaluated. Specifically, the analysis quantifies how active-power oscillations from the load point of interconnection (POI), $P_L$, are attenuated or amplified before reaching the generator POI, $P_H$. This behavior is characterized using the frequency-dependent transfer function $H_{\mathrm{netw}}(f)$:

\begin{equation}
H_{\mathrm{netw}}(f)=
\frac{\Delta P_H(f)}{\Delta P_L(f)},
\label{eq:Hnetw}
\end{equation}

where $\Delta P_L(f)$ and $\Delta P_H(f)$ denote the oscillation amplitudes in $P_L$ and $P_H$, respectively, for a forcing frequency $f$. In general, as the electrical distance between the AI data center and the hydroelectric plant increases, the effective network impedance and damping increase, reducing the magnitude of the oscillatory component transmitted from the load to the generator POI. This attenuation is also frequency dependent, with higher-frequency oscillations generally experiencing stronger attenuation than lower-frequency components. The extent of propagation further depends on network topology. If the generator and data center are radially connected, a larger fraction of the oscillatory disturbance may be transferred along the shared electrical path. In contrast, in a meshed network or a system with multiple feeders and parallel transmission paths, the disturbance is distributed across several branches, such that only a portion of the injected oscillation reaches the hydro-generator POI. Therefore, $|H_{\mathrm{netw}}(f)|$ captures the combined effects of electrical distance, line impedance, damping, frequency, and topology on oscillation propagation.

Next, the susceptibility of the hydro-generator shaft system to oscillations observed at the generator POI is evaluated. The analysis focuses on how oscillations in generator active power propagate through the electromechanical dynamics of the hydropower unit and manifest as cyclic stresses within the turbine-generator shaft system. The relationship is characterized by the frequency-dependent plant transfer function $H_{\mathrm{plant}}(f)$, which relates oscillatory active power $P_H$ at the generator POI (or equivalently the corresponding electromagnetic torque $T_e$) to the resulting shaft torque pulsations $T_{\mathrm{sh}}$:

\begin{equation}
H_{\mathrm{plant}}(f)
=
\frac{\Delta T_{\mathrm{sh}}(f)}
{\Delta P_H(f)}
\approx
\frac{\Delta T_{\mathrm{sh}}(f)}
{\Delta T_e(f)},
\label{eq:Hplant}
\end{equation}

where $\Delta P_H(f)$, $\Delta T_e(f)$, and $\Delta T_{\mathrm{sh}}(f)$ represent the amplitudes of oscillatory generator active power (in p.u.), electromagnetic torque, and mechanical shaft torque responses at forcing frequency $f$, respectively. The response of $H_{\mathrm{plant}}(f)$ depends on both electrical and mechanical characteristics of the unit under study, including generator-to-turbine inertia ratio, shaft stiffness, damping, turbine configuration, unit rating, and inertia distribution across the shaft system. The transfer function provides a direct measure of how strongly oscillatory electrical forcing is amplified by the hydro-generator shaft system. Consequently, transfer-function analysis enables identification of dominant torsional modes, resonance frequencies, and hydro-generator designs or operating conditions associated with elevated susceptibility to AI-induced forced oscillations.\\

\textbf{\textit{Evaluation and Interpretation of Transfer Functions}}: To evaluate the transfer functions $H_{\mathrm{plant}}(f)$ and $H_{\mathrm{netw}}(f)$, this work adopts a frequency-scan methodology in which the system response is evaluated independently at each forcing frequency. A unit-amplitude sinusoidal active-power oscillation is injected at the load POI, and the forcing frequency is swept one value at a time across the selected sub-synchronous frequency range. For each frequency, the steady-state responses in $P_L$, $P_H$, $T_e$, and $T_{\mathrm{sh}}$ are recorded after initial transients have decayed. These responses are detrended to remove steady-state offsets and slow drift, and the observed oscillation amplitudes are used to compute the corresponding transfer-function magnitudes. When the objective is to evaluate only the hydro-generator susceptibility represented by $H_{\mathrm{plant}}(f)$, the forced oscillation may instead be injected directly into $P_H$ at the generator POI. This approach isolates the plant-level torsional response from network propagation effects.

The overall mechanical response of the hydro-generator system to forced oscillations is the combined effect of network propagation and plant-level amplification:
\begin{equation}
\Delta T_{\mathrm{sh}}(f)
\approx
H_{\mathrm{plant}}(f)
H_{\mathrm{netw}}(f)
\Delta P_L(f).
\label{eq:overall_response}
\end{equation}
Peaks in $|H_{\mathrm{plant}}(f)|$ occur when the forcing frequency approaches one of the torsional modes, indicating resonance or near-resonance conditions under which relatively small oscillations in generator active power or electromagnetic torque can produce disproportionately large shaft torque pulsations. The network propagation transfer function $H_{\mathrm{netw}}(f)$ determines how much of the load oscillation reaches the generator output. If a generator is electrically distant from the oscillating data center, the resulting oscillation in generator power output may be small. Hence, even if the oscillation frequency matches a torsional mode and shaft torque oscillations are amplified relative to electromagnetic torque, the total impact on shaft fatigue may remain limited. Therefore, fatigue risk is highest when a significant oscillatory component reaches the generator and its frequency aligns with a torsional mode. Conversely, if the generator is electrically distant from the data center, the transmitted oscillation may be small, limiting the impact even under resonant conditions.

\begin{figure}[t]
    \centering
    \includegraphics[width=0.8\linewidth]{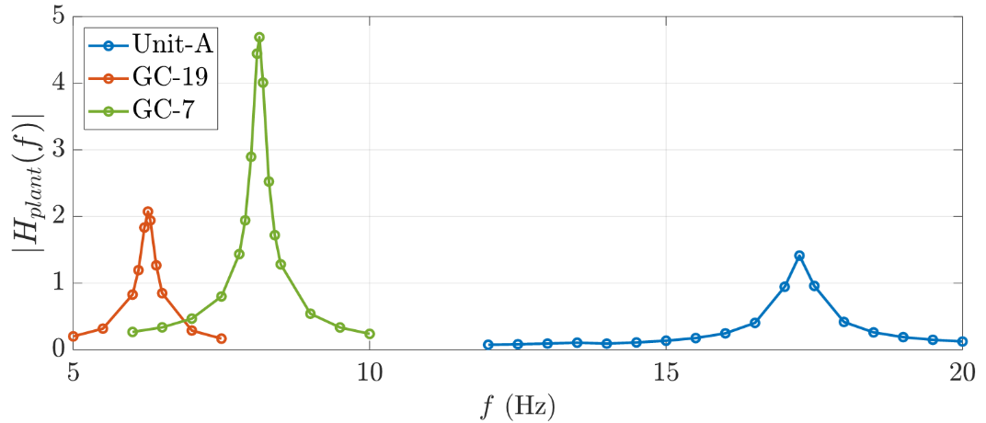}
    \caption{Magnitude of the plant transfer function $H_{\mathrm{plant}}(f)$  for hydro-generator units Unit-A, GC-7, and GC-19.}
    \label{fig:Hplant_response}
\end{figure}

Figure~\ref{fig:Hplant_response} shows the magnitude of the plant transfer function,
$|H_{\mathrm{plant}}(f)|$, computed for the three units listed in Table~\ref{tab:shaft_parameters}.
The transfer functions are evaluated over a frequency range surrounding the respective torsional modes.
As shown in Fig.~\ref{fig:Hplant_response}, the Grand Coulee units GC-7 and GC-19 exhibit maximum torque amplification at torsional mode frequencies of 8.14~Hz and 6.26~Hz, respectively, while Unit A reaches its maximum amplification near 17~Hz. The magnitude of the peak response differs significantly among the units, indicating that torsional amplification is strongly influenced by hydro-generator design parameters. Among the three cases, GC-7 exhibits the largest amplification and therefore represents the highest relative susceptibility to torsional amplification and fatigue under comparable oscillatory forcing conditions. For the Grand Coulee units, the resonant peaks are sharp and confined to a narrow frequency range; that is, the amplification factor decreases rapidly as the forcing frequency moves away from the torsional mode. Therefore, if oscillations introduced by nearby data centers can avoid these narrow frequency regions surrounding the torsional modes, the resulting generator fatigue risk can be substantially reduced. In contrast, the resonance peak for Unit A is comparatively broader, indicating a wider frequency range over which shaft torque amplification may occur.

\section{Sensitivity of Shaft Torque Amplification to Hydro-Generator Parameters}

This section examines how different hydro-generator parameters affect torsional mode frequencies and damping, and consequently, the risk of torsional oscillation amplification. The sensitivity studies are intended to identify which mechanical design features most strongly influence the magnitude and frequency location of the peak shaft torque response.\\

\begin{figure}[h]
    \centering
    \includegraphics[width=0.95\linewidth]{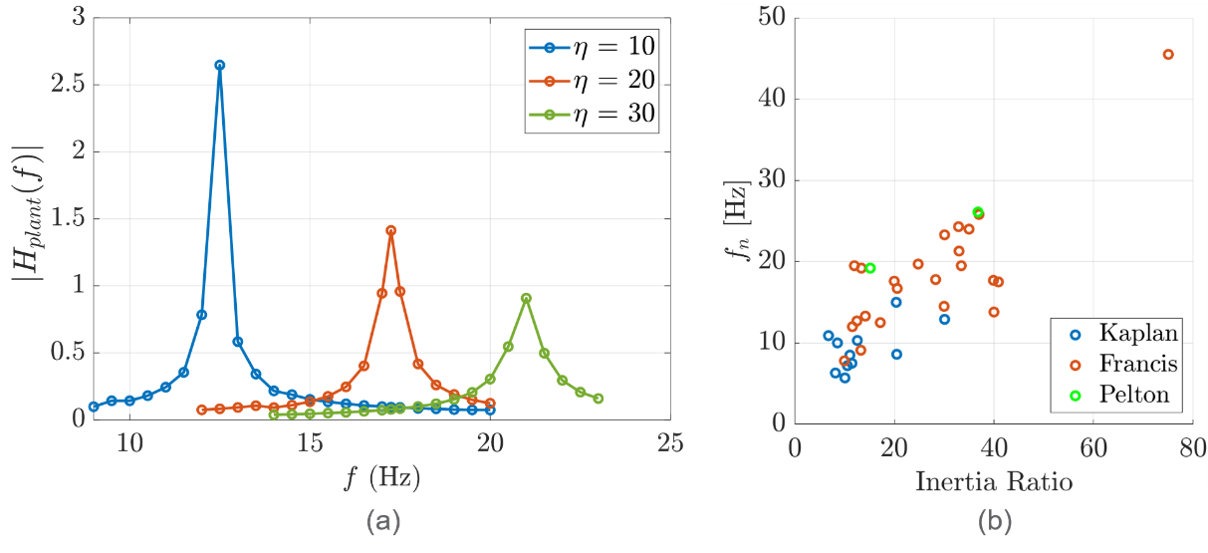}
    \caption{(a) Impact of inertia ratio on $|H_{\mathrm{plant}}(f)|$ evaluated for Unit A and (b) relationship between inertia ratio, turbine type, and torsional mode frequency.}
    \label{fig:inertia_sensitivity}
\end{figure}

\textbf{\textit{Impact of Generator and Turbine Inertia}}: The generator-to-turbine inertia ratio, $\eta$, is defined as the ratio of generator rotational inertia to turbine rotational inertia in the multi-mass shaft system:
\begin{equation}
\eta = \frac{J_g}{J_t},
\label{eq:inertia_ratio}
\end{equation}
where $J_g$ and $J_t$ represent the generator and turbine rotational inertias, respectively. This ratio affects the natural torsional frequency of the turbine-generator shaft mode and therefore influences the forcing frequencies at which torque amplification is most likely. To isolate the effect of inertia distribution, the Unit A parameters listed in Table~\ref{tab:shaft_parameters} are used as the base case. The generator inertia is held fixed at the value provided in Table~\ref{tab:shaft_parameters}, while the turbine inertia is modified to obtain three representative inertia ratios: 10, 20, and 30. For each case, a sinusoidal active-power perturbation is injected at the generator POI, and the forcing frequency is swept across the selected sub-synchronous frequency range. The resulting frequency-response curves are shown in Fig.~\ref{fig:inertia_sensitivity}(a).
The results show that, with all other parameters unchanged, decreasing the inertia ratio shifts the resonant frequency to a lower value and increases the peak torque amplification. To relate these simulated trends to practical hydro-generator designs, Fig.~\ref{fig:inertia_sensitivity}(b) also presents empirical data from~\cite{andersson1984inertia}, where generator-to-turbine inertia ratios for conventional hydroelectric units are reported to commonly fall within the range of 10--30 and vary depending on turbine configuration. The empirical relationship indicates that Kaplan units generally have lower inertia ratios compared with Francis or Pelton units and consequently exhibit lower torsional natural frequencies. The simulation results are consistent with this observed trend: units with lower inertia ratios tend to exhibit lower torsional mode frequencies and greater susceptibility to torque amplification under forced oscillations.

These results suggest that hydroelectric units with lower inertia ratios may experience increased fatigue exposure under persistent sub-synchronous forcing. Since Kaplan units are often associated with lower generator-to-turbine inertia ratios, they may exhibit higher torsional amplification susceptibility compared with comparable Francis or Pelton configurations, depending on shaft stiffness, damping, and specific unit design characteristics.\\

\begin{figure}[h]
    \centering
    \includegraphics[width=0.5\linewidth]{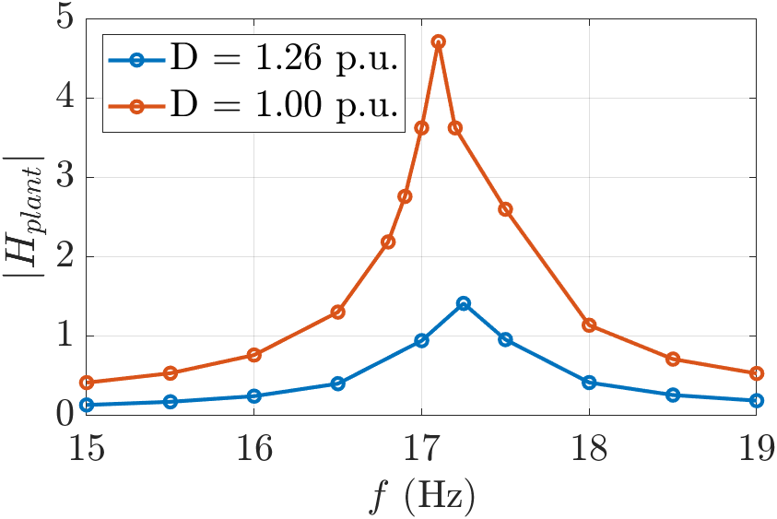}
    \caption{Impact of turbine viscous damping variation on $|H_{\mathrm{plant}}(f)|$ for Unit A.}
    \label{fig:damping_sensitivity}
\end{figure}

\textbf{\textit{Impact of Turbine Viscous Damping}}: The viscous damping provided by water acting on the turbine runner is a key parameter governing shaft torsional amplification. Lower damping reduces the system's ability to dissipate oscillatory energy, increasing mechanical torque pulsations and the associated fatigue risk. Although hydro-generators generally exhibit higher torsional damping than thermal generating units, accurate estimation of turbine damping remains challenging. Hydraulic turbine runners operate within confined water passages, and the effective damping depends on complex fluid--structure interactions, turbine geometry, operating point, and water-column dynamics, all of which are difficult to represent accurately in simulation models. To account for this uncertainty, torsional amplification is evaluated over a range of damping values below the nominal estimate to establish conservative worst-case risk bounds. The degree of conservatism depends on the selected damping range. In this study, the turbine damping parameter for Unit A is varied from the baseline value of 1.26 p.u. (Table~\ref{tab:shaft_parameters}) down to 1.0 p.u., as shown in Fig.~\ref{fig:damping_sensitivity}. The results demonstrate strong sensitivity to the damping parameter: even a modest reduction in damping produces a substantial increase in torsional amplification near resonance.

\section{Assessment of Shaft Fatigue from Torque Amplification}

This section describes how sustained oscillations in the shaft torque of a turbine-generator, excited by a load oscillating at the unit's torsional frequency, can be translated into a shaft-fatigue assessment. First, two standard tools for high-cycle fatigue analysis, the S--N curve and Goodman diagram, are reviewed. Fatigue assessment based on torque amplification obtained from the EMT simulations described previously is then demonstrated using the two Grand Coulee unit examples listed in Table~\ref{tab:shaft_parameters}. Accurate fatigue analysis is highly complex and depends on plant-specific configuration parameters. The illustrative example presented here is intended to demonstrate the methodology; therefore, the results should not be directly interpreted as plant-specific conclusions.

In the two-mass torsional representation used in this paper, the turbine runner and generator rotor are modeled as rotating inertias connected by an elastic shaft. When the two masses move relative to one another, the shaft transmits an internal torque proportional to the angular twist between them. This transmitted shaft torque represents the mechanically relevant loading quantity for fatigue assessment. For a hollow circular shaft with outer diameter $D$ and inner diameter $d$ transmitting a torque $T$, the surface shear stress $\tau$ is given by
\begin{equation}
\tau = \frac{T(D/2)}{J},
\end{equation}
where the polar second moment of area is
\begin{equation}
J=\frac{\pi}{32}(D^4-d^4).
\label{eq:polar_moment}
\end{equation}
A torque oscillation of amplitude $\Delta T_{\mathrm{sh}}$ about a mean torque $\bar{T}_{\mathrm{sh}}$ produces an alternating shear stress
\begin{equation}
\tau_a=
\frac{\Delta T_{\mathrm{sh}}(D/2)}{J},
\label{eq:alternating_stress}
\end{equation}
superimposed on a mean shear stress
\begin{equation}
\tau_m=
\frac{\bar{T}_{\mathrm{sh}}(D/2)}{J}.
\label{eq:mean_stress}
\end{equation}
The resulting stress pair $(\tau_m,\tau_a)$ governs the high-cycle fatigue behavior of the shaft.\\

\textbf{\textit{S--N Curves}}: The fatigue resistance of a material is characterized by its S--N curve, which relates stress amplitude $S$ to cycles to failure $N$. These curves are typically obtained from rotating-bending or axial fatigue tests~\cite{asm_fatigue,shigley2025}. For ferrous alloys such as forged shaft steels, the S--N curve approaches a horizontal asymptote. Below a threshold stress amplitude, known as the endurance limit ($S_e$), the material can sustain effectively unlimited cycles without fatigue failure. A common estimate for the laboratory endurance limit of steel is
\begin{equation}
S'_e \approx \min(0.5S_u,290~\mathrm{MPa}),
\label{eq:endurance_limit}
\end{equation}
where $S_u$ is the ultimate tensile strength~\cite{asm_fatigue,shigley2025}.
This polished-specimen endurance limit is corrected for the actual component using Marin factors:
\begin{equation}
S_e=k_a k_b k_e S'_e,
\label{eq:marin_factor}
\end{equation}
where $k_a$, $k_b$, and $k_e$ account for surface finish, component size, and reliability, respectively. For large hydro-generator shafts, these correction factors can substantially reduce the endurance limit. Typical forged surfaces may result in $k_a \approx 0.4$--$0.5$, while large shaft diameters may result in $k_b \approx 0.5$--$0.6$. Because the loading considered here is torsional, the bending endurance limit is converted to a shear endurance limit using
\begin{equation}
S_{se}\approx0.577S_e.
\label{eq:shear_endurance}
\end{equation}

\par \textbf{\textit{Goodman Diagram}}: An S--N curve characterizes fully reversed loading with zero mean stress. However, a generator shaft operates under a large steady torque associated with normal power transfer. Therefore, the torsional oscillation introduces an alternating stress component superimposed on a mean stress. The Goodman diagram captures the detrimental effect of mean stress on fatigue life~\cite{barsom1999fracture,shigley2025}. The mean stress is plotted on the horizontal axis and alternating stress on the vertical axis. The Goodman failure line connects the shear endurance limit $S_{se}$ on the alternating-stress axis to the ultimate shear strength $S_{su}$ on the mean-stress axis.
Operating points below the Goodman line are considered acceptable for high-cycle fatigue under the assumed material model, whereas points on or above the line indicate reduced fatigue margin or finite-life risk. The yield boundary represents stress combinations that cause permanent deformation during a single loading cycle.
\begin{figure}[h]
    \centering
    \includegraphics[width=0.65\linewidth]{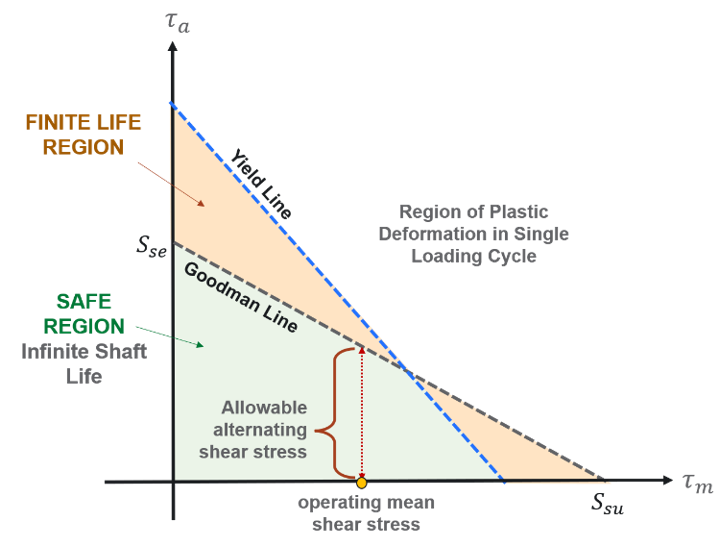}
    \caption{The Goodman diagram. The line from the endurance limit to the ultimate strength separates the safe (infinite-life) region from finite-life/failure. Mean stress erodes the tolerable alternating stress: a point with high mean stress fails at a lower alternating amplitude than one near the origin.}
    \label{fig:placeholder}
\end{figure}

The fatigue safety factor $n$ for an operating point $(\tau_m,\tau_a)$ is obtained using the modified Goodman relation~\cite{shigley2025}:

\begin{equation}
\frac{1}{n}
=
\frac{\tau_a}{S_{se}}
+
\frac{\tau_m}{S_{su}} .
\label{eq:goodman}
\end{equation}

This relation provides a direct connection between the torque-amplification metric and shaft fatigue assessment. Values of $n>1$ indicate operation within the safe region, while lower values indicate reduced fatigue margin. Values of $n<1$ indicate that the estimated stress state exceeds the high-cycle fatigue limit.\\

\begin{figure}[h]
\centering
\includegraphics[width=\linewidth]{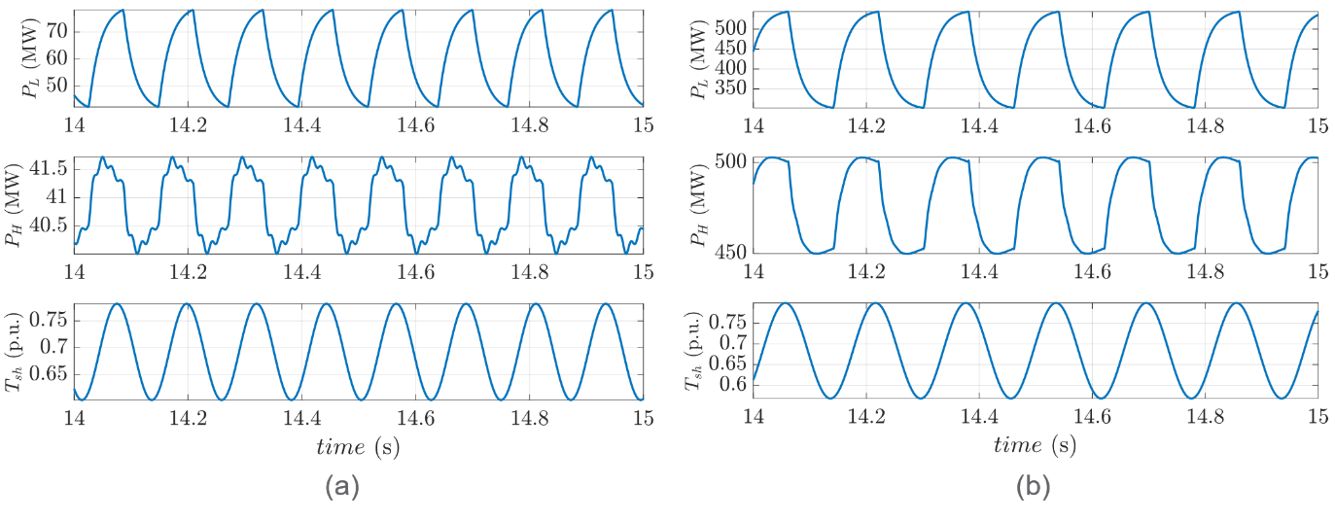}
\caption{(a) GC-7 unit with 8.14~Hz power oscillation and (b) GC-19 unit with 6.26~Hz power oscillation, both operating at approximately 80\% mean loading level.}
\label{fig:GC_response}
\end{figure}

\textbf{\textit{Fatigue Assessment for GC-7 and GC-19 Units}}: To demonstrate the fatigue assessment methodology, EMT simulation results are evaluated for the Grand Coulee unit examples listed in Table~\ref{tab:shaft_parameters}. GC-7 represents a smaller 60~MVA machine, whereas GC-19 represents a large 715~MVA generator. Scenarios are simulated in which each unit operates at approximately 80\% of rated active power while the AI load oscillates at the corresponding shaft torsional frequency, as shown in Fig.~\ref{fig:GC_response}. The AI load demand and oscillation amplitudes are scaled according to the generator ratings. The generator terminal oscillations are approximately 3~MW (5\% of MVA rating) and 50~MW (6\% of MVA rating) for GC-7 and GC-19, respectively. The mean and alternating shaft torques obtained from EMT simulations are converted into nominal shear stresses using (3). Based on shaft geometries reported in~\cite{eilts1979shaft}, the resulting stresses are summarized in Table~\ref{tab:fatigue_analysis}. The actual shaft materials are not publicly available; therefore, representative properties for forged steel are assumed.

Substituting the calculated stresses into (4) gives fatigue safety factors of $n=5.9$ and $n=3.8$ for GC-7 and GC-19, respectively, as shown in Table~\ref{tab:fatigue_analysis}. Both operating points remain within the Goodman safe region, indicating that the simulated torsional oscillations, under the assumed material properties and damping conditions, do not introduce immediate fatigue concerns.
\begin{table}[h]
\centering
\caption{Fatigue analysis of Grand Coulee units GC-7 and GC-19.}
\label{tab:fatigue_analysis}
\renewcommand{\arraystretch}{1.15}
\begin{tabular}{lcc}
\toprule
\textbf{Parameter} & \textbf{GC-7} & \textbf{GC-19}\\
\midrule
Mean loading level & 80\% & 80\%\\
Shaft diameters $(D,d)$ & 
686 mm, 160 mm &
2540 mm, 2134 mm\\
Mean shaft torque $\bar{T}_{\mathrm{sh}}$ &
0.692 p.u. &
0.684 p.u.\\
Mean shear stress $\tau_m$ &
31.4 MPa &
40.2 MPa\\
Alternating torque $\Delta T_{\mathrm{sh}}$ &
0.089 p.u. &
0.116 p.u.\\
Alternating shear stress $\tau_a$ &
4.05 MPa &
6.8 MPa\\
Fatigue safety factor $n$ &
5.9 &
3.8\\
\bottomrule
\end{tabular}
\end{table}
The example simulations show that 5--6\% peak-to-peak active power oscillations occurring while a unit operates at approximately 80\% rated power remain below the high-cycle fatigue limit for representative large and small hydro units. The inherently higher torsional damping of hydro units compared with steam and natural gas turbines may therefore provide an advantage by reducing fatigue accumulation from oscillatory interactions associated with nearby AI data center loads. This observation does not imply that persistent forcing can be ignored. Instead, the risk can be managed through periodic inspections and condition monitoring to identify crack initiation or propagation at shaft discontinuities and other fatigue-prone locations. When combined with appropriate monitoring strategies, hydro units may provide additional flexibility for accommodating large data center clusters by tolerating higher power fluctuations in the torsional frequency range. This may reduce the need for extensive rack-side power management strategies and additional load smoothing solutions such as battery energy storage systems with advanced control schemes.

\section{Conclusions}
This paper presents a model-based framework for assessing hydro-generator shaft-fatigue risk from AI data center load oscillations, providing a starting point for evaluating an emerging and currently under-studied source of mechanical stress on hydroelectric assets. The proposed methodology can support data center interconnection studies and asset risk screening as hyperscale loads continue to expand in hydropower-rich regions. Analysis using representative plant models highlights some key takeaways. 

First, fatigue risk is the product of two largely independent effects – how much oscillation reaches a generator (network propagation), and how strongly the unit’s shaft system amplifies what arrives (plant susceptibility). Practically, this means a data center oscillating exactly at a plant's torsional frequency may pose little risk if the plant is electrically remote, while a plant close to a large, persistent oscillation source can face exposure even off resonance. The transfer-function approach proposed here gives grid operators, data center developers, and plant owners a practical way to flag when detailed site-specific analysis is warranted, and a shared basis for discussing which frequency ranges to avoid and when oscillatory behavior calls for remediation.

Second, plant-level susceptibility is governed primarily by generator-to-turbine inertia ratio and effective damping. Controlled sensitivity sweeps show that lower inertia ratios shift the dominant torsional mode to lower frequency and increase peak amplification, while lower damping sharpens and raises the resonant response. Because published data indicates that Kaplan units tend toward lower ratios than comparable Francis or Pelton units, they warrant closer attention — though actual risk also depends on each unit’s stiffness and damping, not the turbine type alone. Since it is difficult to obtain exact damping values at a plant, risk screening can use a plausible range of damping values rather than a single estimate, supplemented where possible by disturbance-based estimation of torsional mode properties.

Third, translating simulated shaft torque amplitudes into a Goodman safety factor gives a single screening-level risk metric. While this simplified methodology cannot replace detailed plant-specific fatigue studies, it can flag cases where they are needed. Since hydro units generally have higher torsional damping than thermal plants, their fatigue exposure to AI-workload-induced persistent forcing may be lower. Understanding this risk can help define monitoring and mitigation actions needed to limit impacts to plant operational life while supporting large data center integration.

Several open issues identified during this work warrant further study. The Grand Coulee comparison in this paper shows that torsional frequencies can vary meaningfully even among units within the same plant – raising the question of how performance requirements for nearby data centers should account for that variation. In future work, the authors aim to extend the EMT simulations to study the impact on multiple units within a hydroelectric power plant, the impact of multiple hyperscale sites in a hydro-rich cluster, and the accumulation of fatigue damage under realistic, non-stationary AI load profiles using rainflow-counting methods rather than single-frequency forcing alone.

\section*{Acknowledgments}
This work was supported by the U.S. Department of Energy's Office of Critical Minerals and Energy Innovation through the Hydropower and Hydrokinetic Office as part of the HydroWIRES Initiative. The authors also acknowledge valuable technical discussions and inputs from Dmitry Kosterev of Bonneville Power Administration (BPA), Shawn Patterson of the U.S. Bureau of Reclamation (USBR), and Sean Brosig of the U.S. Army Corps of Engineers (USACE).

\bibliographystyle{IEEEtran}
\bibliography{main}
\end{document}